\documentclass[fleqn,usenatbib]{mnras}

\usepackage{newtxtext,newtxmath,verbatim}

\usepackage[T1]{fontenc}
\usepackage{ae,aecompl}
\usepackage{footmisc}
\newcommand{\defaultfirerun}{\textbf{m10q.FIRE-2}}
\newcommand{\nugridrun}{\textbf{m10q.nugrid}}
\newcommand{\nugridavgrun}{\textbf{m10q.nugrid\_avg}}

\usepackage{graphicx}	

\usepackage{amsmath}	
\usepackage{amssymb}	

\newcommand{\mstar}{M_{\rm \star}}
\newcommand{\msun}{\rm \, M_{\odot}}

\title[Progenitor-dependent metal yields]{Progenitor-mass-dependent yields amplify intrinsic scatter in dwarf-galaxy elemental abundance ratios}

\author[D. Muley et al.]{
Dhruv A. Muley$^{1, 2}$\thanks{E-mail: dmuley@berkeley.edu},
Coral R. Wheeler$^{3, 4, 5}$\thanks{Hubble Fellow},
Philip F. Hopkins $^{3}$,
Andrew Wetzel $^{6}$, \newauthor
Andrew Emerick $^{3, 5}$,
Du{\v s}an Kere{\v s}$^{4}$
\\
$^{1}$Department of Astronomy, University of California, Berkeley\\
$^{2}$Department of Physics and Astronomy, University of Victoria\\
$^{3}$California Institute of Technology, Pasadena\\
$^{4}$University of California, San Diego\\
$^{5}$Carnegie Observatories, Pasadena\\
$^{6}$Department of Physics and Astronomy, University of California, Davis
}

\date{Accepted XXX. Received YYY; in original form ZZZ}

\pubyear{2020}

\begin{document}
\label{firstpage}
\pagerange{\pageref{firstpage}--\pageref{lastpage}}
\maketitle

\begin{abstract}

We explore the effect of including progenitor mass- and metallicity-dependent yields, supernova rates and energetics on variations in elemental abundance ratios (particularly [$\alpha$/Fe]) in dwarf galaxies. To understand how the scatter and overall trends in [$\alpha$/Fe] are affected by including variable metal yields from a discretely sampled IMF, we run FIRE simulations of a dwarf galaxy $(\mstar($z = 0$) \sim 10^6\msun)$ using nucleosynthetic yields from the \texttt{NuGrid} database that depend on the stellar progenitor mass and metallicity. While \texttt{NuGrid} exhibits lower aggregate $\alpha$-element production than default-FIRE yields, we find that its explicit mass dependence, even when including turbulent metal diffusion, substantially widens the intrinsic scatter in the simulated [Fe/H]-[$\alpha$/Fe] -- a phenomenon visible in some observations of dwarf galaxies.

\end{abstract}




\section{Introduction}

Recent detailed observations of dwarf galaxies have helped constrain their elemental evolution and star-formation histories with ever-increasing precision. Abundance measurements from medium-resolution spectroscopy can detect a variety of different species in galaxies, including Fe, Mg, Si, Ca and Ti. These measurements have been used to estimate the overall stellar metallicity and metallicity distributions for Milky Way dwarf spheroidal (dSph) satellite galaxies \citep{Shetrone2001, Cohen2010, Tafelmeyer2010, Venn2012,  Starkenburg2013}, 
ultra-faint dSphs \citep{Frebel2010,Vargas2013}, and for satellites of M31 \citep{Vargas2014,Kirby19,Wojno20}. These observations have revealed, among other features, a tight stellar mass-metallicity relation (MZR) spanning five orders of magnitude in stellar mass \citep{Kirby11,Kirby13,Ho2014}, narrow metallicity distributions \citep{Kirby2010}, and a generally decreasing trend in [$\alpha$/Fe] versus [Fe/H] \citep{Hasselquist2017, Hill2019, Kirby09, Kirby2010, Letarte2010, Tolstoy09}.

Comparing these high-quality observations to one-zone chemical evolution models \citep[e.g.,][]{Lanfranchi03,Cescutti2008, Kirby11, Kirby11b} as well as hydrodynamical simulations, has improved our understanding of timescales and yields for processes such as AGB winds, Type Ia supernovae, and Type II core-collapse supernovae, but some questions still remain. While many observational studies of dwarfs have found that they have a narrow scatter in $\alpha$ abundances  \citep[see e.g.][for a compilation]{Tolstoy09,Revaz2016}, observations of some dwarfs may indicate a much greater intrinsic scatter \citep{Aoki2009, Lemasle2012}, particularly at low metallicities/early times when Type II supernovae dominate enrichment \citep{Frebel15, Vargas2013, Vargas2014}.
The connection between progenitor mass, metallicity, and $\alpha$-element yields from Type II supernovae has long been known \citep{Woosley1995,Tsujimoto1995}, and a number of simulated works have incorporated some form of progenitor mass- and metallicity-dependent yields into their chemical enrichment models,\cite{Mosconi2001,Kawata2006,Stinson2006,Wiersma2009,Schaye2015}. However, few have done so with the resolution required to investigate dwarf galaxies -- particularly with time resolution sufficient to allow for the mixing of the varying yields within a dynamical time (\citealt{Few2012,Pillepich2018}, but see \citealt{Revaz2018,Jeon2017} for high resolution examples). Fewer still (see \citealt{Emerick2019} for an exception) have specifically investigated how scatter in the [$\alpha$/Fe] versus [Fe/H] relationship is affected by averaging yields from Type II supernovae and stellar winds (if included) over the initial mass function \citep[IMF; ][]{Valcke2008,Okamoto2010}, as is still done in a number of simulations \citep{Marcolini2008,Schroyen2013, Wang2015,Dave2016,Starkenburg2017,Hirai2017,Dave2019}. This includes the standard FIRE-2 simulations from \citet{Hopkins2018} to which we directly compare in this work. One notable exception is \citet{Revaz2016}, who use non-cosmological simulations of dwarf galaxies to investigate the effect that high resolution has on artificially increasing the scatter in [$\alpha$/Fe] versus [Fe/H] due to incomplete sampling of the IMF. They find that the scatter is reduced when a metal-mixing scheme is included, but do not explicitly compare to the case in which the yields are averaged over the IMF.


\begin{figure*}
    \centering
    \includegraphics[width=1\textwidth]{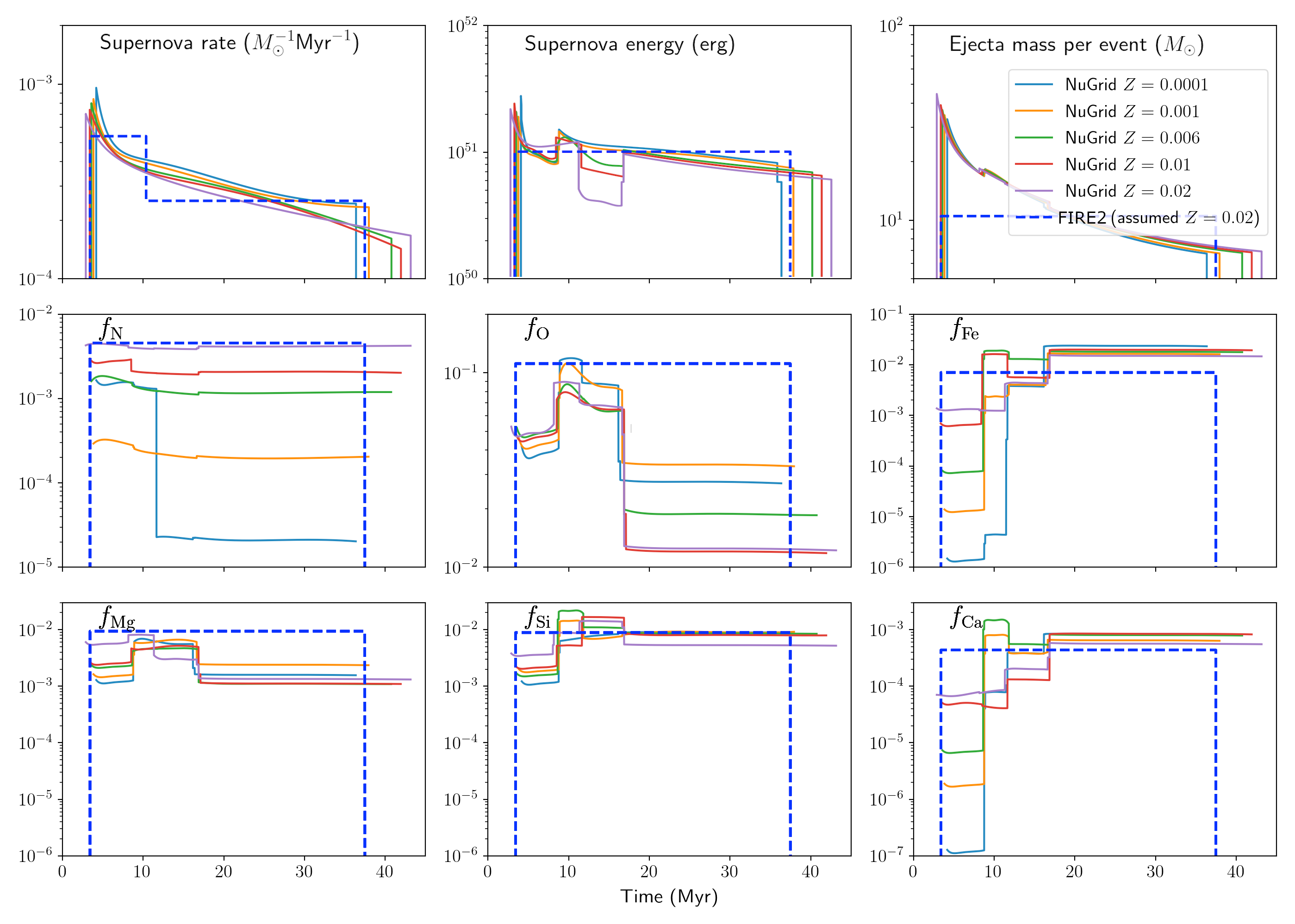}
    \caption{\textit{Upper row}, aggregate properties for Type II core-collapse supernovae in \texttt{NuGrid} at the five metallicities it samples. Supernova rate is expressed in events per Myr per $\msun$ in a star particle, as a function of star particle age. \textit{Lower two rows} show mass fractions of selected metals in core-collapse ejecta, as a function of time. For most progenitor masses, N exhibits a roughly linear metallicity dependence (as it does in the pre-existing FIRE-2 treatment, though this is not shown in the figure); in other cases, the dependence is usually moderate in strength and non-monotonic.
    } 
    \label{fig:sb99_nugrid_full}
\end{figure*}

In this paper, we present a case-study of the effect of IMF averaging on [$\alpha$/Fe] vs [Fe/H] by re-simulating the $M_{\star} \sim 10^6 M_\odot$ \textbf{m10q} dwarf galaxy from the FIRE-2 project \citep{Hopkins2018} with both IMF-averaged and properly progenitor mass- and metallicity-dependent yields. We modify the yield, SNe rate, and energy models in three different runs: the default, IMF-averaged, metallicity-independent FIRE-2 yields \citep[\defaultfirerun, based on][]{Leitherer1999, Nomoto2006, Hopkins2018}, the progenitor-mass- (and metallicity-) dependent \texttt{NuGrid} database of post-processed stellar evolution simulations \citep[\nugridrun, based on][]{Pignatari2016, Ritter2018, Ritter2018b}, and IMF-averaged \texttt{NuGrid} yields (\nugridavgrun). We show that progenitor-mass dependence (but not metallicity-dependence alone) raises the \textit{scatter} in [$\alpha$/Fe]-[Fe/H] to levels commensurate with some observations, but that neither changing the yields alone, nor adding in the mass and metallicity dependencies, have a dynamically significant effect on the simulated galaxy. However, we do find that, at low [Fe/H], using \texttt{NuGrid} yields results in a systematically lower [$\alpha$/Fe] abundance than is observed in many (but not all) dwarf galaxies, implying that its treatment of some yields (particularly Mg) may need to be revisited.

\section{Methods}
\label{sec:methods}
In line with existing FIRE simulations, we use the meshless finite-mass (MFM) hydrodynamical method \citep{Hopkins2015} at a resolution of $m_{\rm bar}  = 250 \msun$ \cite[e.g.,][]{Wheeler19}. We track 11 species: H, He, C, N, O, Ne, Mg, Si, S, Ca, and Fe; and three enrichment processes: Type II core-collapse supernovae, O/AGB winds, and Type Ia supernovae\citep[][]{Hopkins2018}. Wind yields and energies are deposited continuously, while supernovae are drawn each timestep from a binomial distribution with probability $p$ based on the event rate, timestep duration, and star particle mass; when an event occurs, we assign its yields and energy as a function of star-particle age. For \nugridrun, we use the \texttt{SYGMA} \citep{Ritter2018} code to construct time- and metallicity-dependent supernova yields and event rates for a \cite{Kroupa2002} stellar population, whereas for \nugridavgrun, we average these yields over the IMF. In all cases, the IMF runs from $0.1-100\msun$, with a Type II supernova progenitor minimum mass of $8\msun$. As a check of the effect of this lower mass cutoff, we created a sample stellar population with an assumed 5-100 Msun range of supernova progenitors, and found that the minimum and maximum [alpha/Fe] values in supernova ejecta remain the same, and so will not significantly affect the results presented here. On average, \texttt{NuGrid} core-collapse supernovae produce ${\sim}13 M_\odot$ of ejecta per event, while the default FIRE-2 core-collapse supernovae produce 10.5 $M_\odot$. Given the timesteps we use, $p$ never exceeds 1, so our supernovae are ``time-resolved'' (i.e., we avoid over- or undercounting). This in turn ensures that on average, \nugridrun~accurately reproduces the yield profile from the tables.

Because \texttt{NuGrid}'s AGB yields assume solar abundance ratios, we use them only for the total mass-loss rate, but then partition the ejecta between elements according to their surface abundances (as in \citealt{Hopkins2018}). We employ the same treatment in \nugridavgrun, after IMF-averaging to eliminate time dependence. For type Ia supernovae in both the \nugridrun \ and \nugridavgrun \ simulations, we continue to use the pre-existing \cite{Hopkins2018} yields, but with a simple $R_{\rm Type \ Ia}\propto t^{-1.1}$ delay-time distribution \citep[from][]{Maoz2017} rather than the Gaussian prompt+constant delay treatment \citep{Mannucci2006} used in \defaultfirerun~and in previous FIRE-2 simulations. This increases the number of Type Ias by ${\sim}2\times$, to ${\sim}15\%$ of all supernovae, but does not otherwise affect the main results presented here.

\label{sec:typeII}
\begin{figure}
    \centering
    \includegraphics[trim={0 0 0 1.2cm},clip, width=0.5\textwidth]{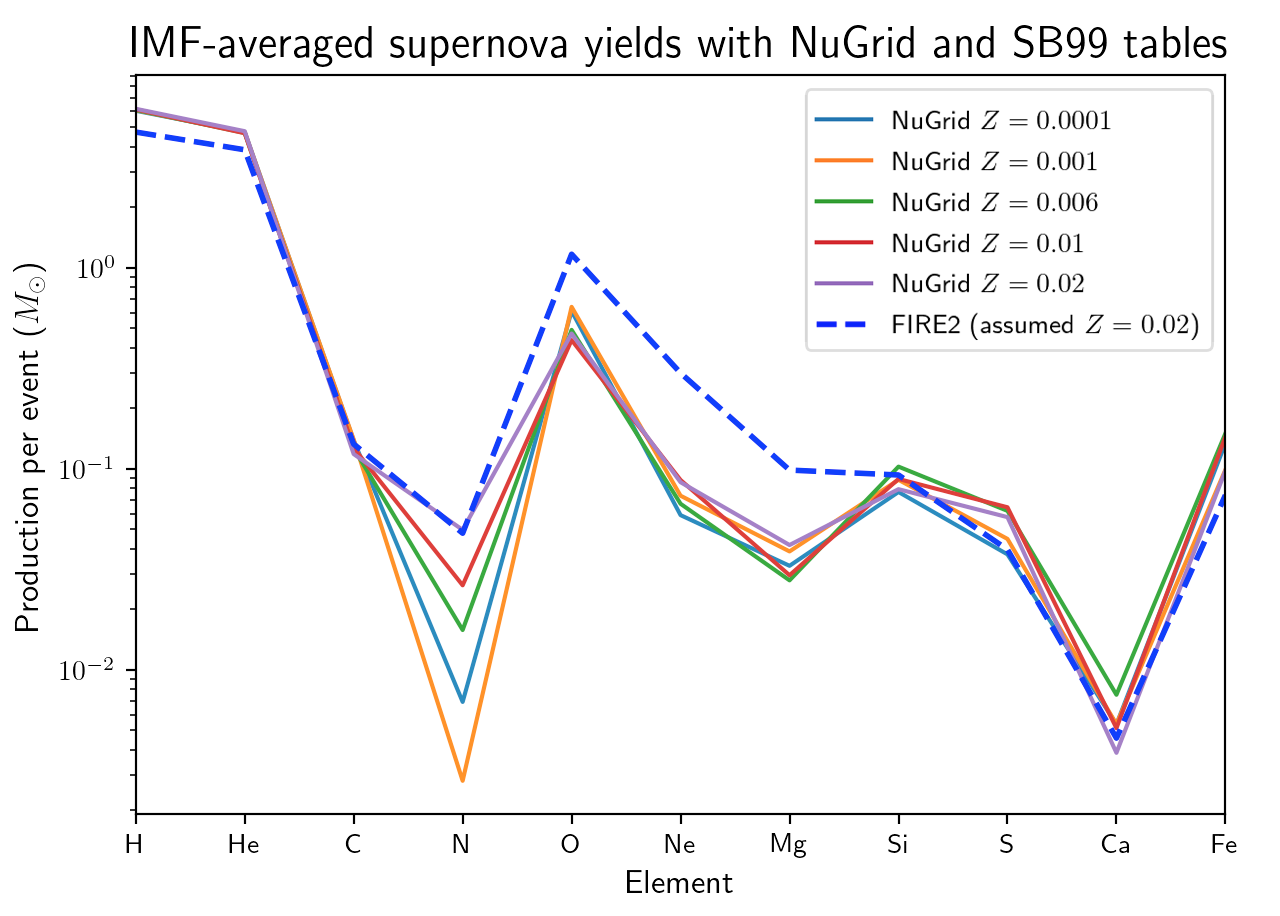}
    \caption{\textit{Above}, the IMF-averaged Type II core-collapse supernova yields of the FIRE-2 default and \texttt{NuGrid}-based yield prescriptions. While most yields are similar, production of O, Ne, and Mg differs substantially between the models. IMF-averaged metallicity dependence in \texttt{NuGrid} is weak, except for production of N (which scales linearly with metallicity in default FIRE-2, although for clarity we omit this on the plot).
    }
    \label{fig:sb99_nugrid_imf_avg}
\end{figure}
In the top row of Figure \ref{fig:sb99_nugrid_full}, we plot the rates, energies, and ejecta mass for core-collapse supernova events in \texttt{NuGrid}, alongside the \texttt{STARBURST99}-based \citep{Leitherer1999} FIRE-2 default fits. As expected, the quantities obtained from \texttt{NuGrid} decline with time, due to the decreasing progenitor mass, with only a weak dependence on metallicity. In an integrated sense, however, the time-integrated difference in supernovae per solar mass formed between the two models is ${\sim}20\%$, and differences in rate at any given time are no more than a factor of several. This implies (and our simulations bear out) that for galaxy-averaged properties such as star-formation rates and total metallicities, there should be no qualitatively meaningful difference between yield models.

In the second and third rows, we plot the ejecta mass fraction of various elements as a function of age (equivalently, progenitor mass) from the \texttt{NuGrid} tables, along with the IMF-averaged default FIRE-2 yields from \cite{Nomoto2006}. For O, we find a high mass-fraction in ejecta from the most massive progenitors, followed by a metallicity-dependent drop-off; for N, production scales linearly in metallicity ($f_{\rm N} \approx 4.05 \times 10^{-3} (Z/Z_\odot)$), with the exception of massive progenitors at the lowest metallicity $Z = 10^{-4}$.

Of the four $\alpha$-elements comprising the \cite{Kirby11} observational definition, we directly track Mg, Si, and Ca, plotted in the bottom row of Figure \ref{fig:sb99_nugrid_full}. Mass fractions for these elements, along with Fe, exhibit sharp jumps at high progenitor mass, before stabilizing at later times. The times and magnitudes of these jumps differ for different metallicities (and the metallicity-dependence differs for each element), so the resulting [(Mg, Si, Ca)/Fe] of ejecta can vary by several dex between different core-collapse events from the same star particle. To facilitate comparison with default FIRE-2, we do not directly simulate Ti, but rather use Ca as a tracer in post-processing (estimating ${\rm [}\alpha{\rm /Fe]} \equiv (1/4)({\rm [Mg/Fe] + [Si/Fe]} + {\rm [Ca/Fe]} + {\rm [Ti/Fe]}) \approx (1/4)({\rm [Mg/Fe] + [Si/Fe]} + 2{\rm [Ca/Fe]})$). We motivate this choice by the fact that in the \texttt{NuGrid} tables, [$\alpha$/Fe] for core-collapse ejecta almost\footnote{\label{fn:rare_higher_dex}Core-collapse supernovae in 8-12 Myr old star particles (accounting for ${\sim}7\%$ of events), specifically at $Z = 10^{-4} {\rm \ or \ } 10^{-2}$, show a difference between defined and estimated [$\alpha$/Fe] of ${\sim}0.3 {\rm \ dex}$. This is still substantially smaller than the range in [$\alpha$/Fe] of a star particle's core-collapse ejecta over all time, which varies from 0.88 to 3.00 dex (0.91 to 3.34 dex in our estimate), with higher values associated with lower metallicities.} always differs by ${\lesssim}0.04$ dex between definition and estimate for individual supernovae, and \textit{always} by ${\lesssim}0.03$ dex in IMF-average---much smaller than errors from interpolation or the yield model itself. In any case, the specific estimate we use, or even individual abundance ratios, does not meaningfully impact our results.

As a test of consistency, we plot in Figure \ref{fig:sb99_nugrid_imf_avg} the IMF-averaged core-collapse supernova yields from \texttt{NuGrid} versus the pre-existing FIRE-2 yields \citep[based on][]{Nomoto2006}. For almost all elements (except N) we find that IMF-averaging largely suppresses metallicity dependence; overall, a FIRE-2 default supernova produces 70\% more metals than one with \texttt{NuGrid}-based yields. Yields for H, He, C, Si, S, and Ca, agree between the two models to within ${\sim}20\%$. For other elements the discrepancies are more pronounced: O production is reduced in \texttt{NuGrid} by a factor of ${\sim}2-3$ depending on metallicity, while that of Ne is smaller by a factor of ${\sim}3-5$. Mg yields are ${\sim}2-4\times$ lower, while those for Fe is larger by a factor of ${\sim}1-2$. Consequently, in an IMF-averaged sense, the [$\alpha$/Fe] of core-collapse supernova ejecta is lower by ${\sim}0.3$ dex in \texttt{NuGrid} with respect to FIRE-2-default models. While these differences in normalization impact overall trends in galactic elemental abundances, they do not change the increased \textit{dispersion} in [$\alpha$/Fe] due to progenitor-mass dependence.

The MFM hydrodynamics method used here has no mass fluxes between fluid elements, so without additional modeling, metals deposited into a gas element would be locked there permanently, not mixing with surrounding gas elements. As a consequence, random variations in metal enrichment due to particle configuration, or which star particles have supernovae of a given mass when, would add to create substantial noise in a galaxy's [$\alpha$/Fe] and [Fe/H] distributions. This is unphysical; realistically, marginally-unresolved fluid turbulence and instability should exchange metals between adjacent fluid elements and suppress this noise. Therefore, FIRE-2 simulations explicitly model sub-grid metal diffusion/mixing \citep{Hopkins2017}. \cite{Escala18} showed that for MFM, metal diffusion is required to eliminate noise and make detailed comparisons between simulated and observed [$\alpha$/Fe] and [Fe/H] in galaxies; their results are robust with respect to the specific choice of diffusion coefficient. \cite{Williamson2016} obtain similar results for metal diffusion in smoothed-particle hydrodynamics (SPH) simulations, where there are also no fluxes between fluid elements. These properties are especially important for the simulations we present here, due to substantial variation in the [$\alpha$/Fe] of core-collapse ejecta from different-mass progenitors \footref{fn:rare_higher_dex}.

\section{Results}
\label{sec:results}
\subsection{Star-formation rates}
\label{sec:sfh}

\begin{figure}
    \centering
    \includegraphics[width=0.469\textwidth]{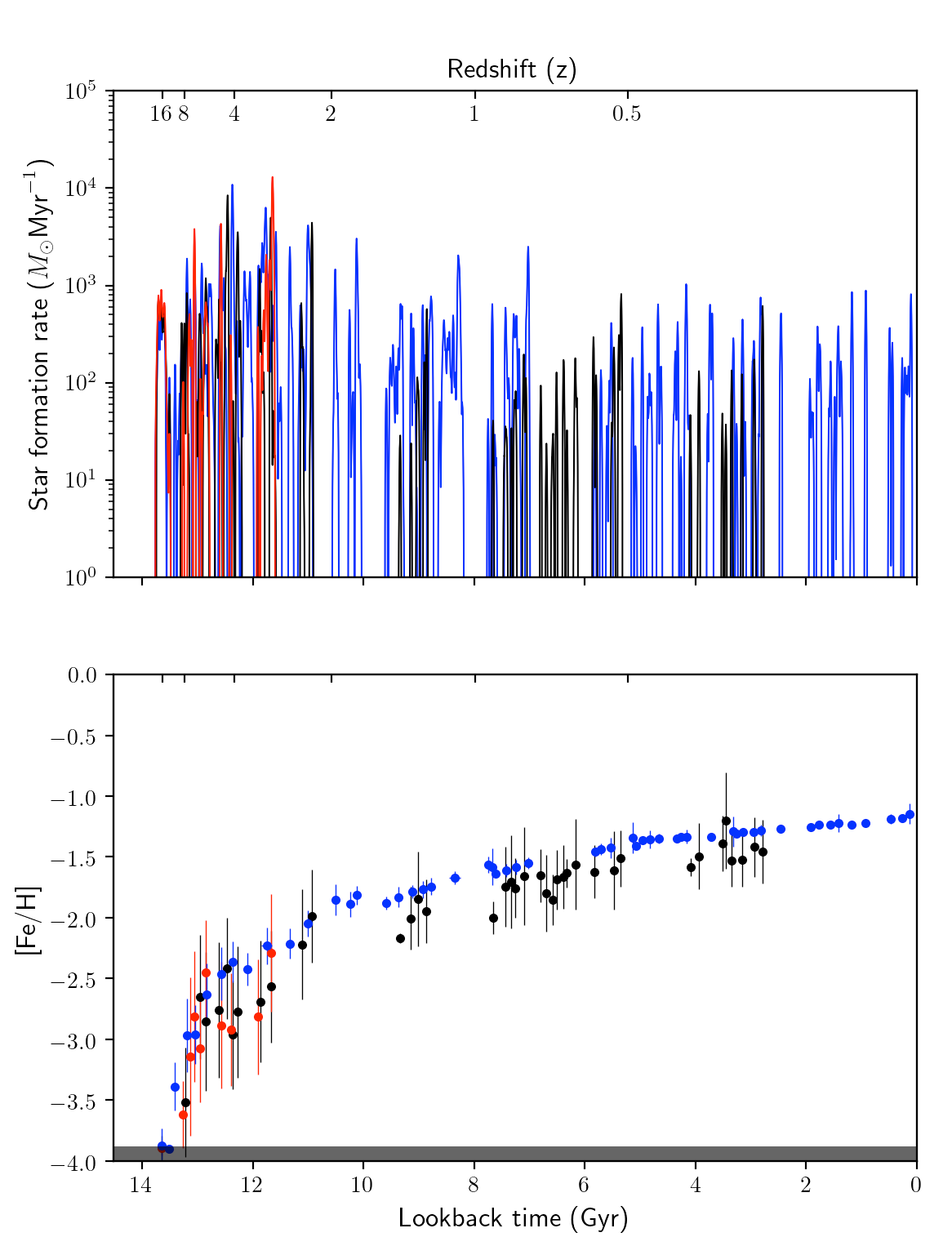}
    \includegraphics[width=0.465\textwidth]{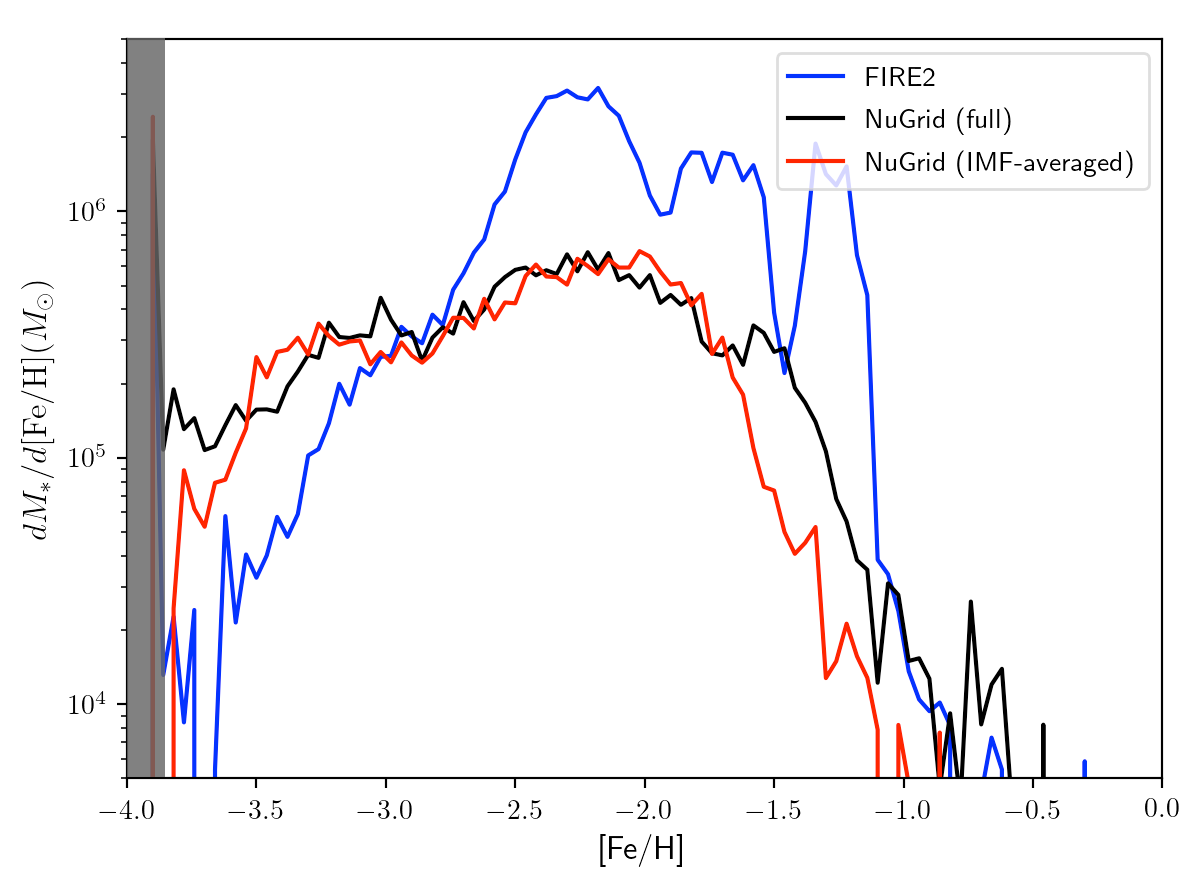}
    \caption{\textit{Above}, the star-formation rate in each of our simulations over time, smoothed on a 20 Myr timescale; the \texttt{NuGrid} runs stochastically self-quench by $z \sim 2$, as in default-FIRE-2 runs \citep[e.g.,][]{Su2018}. \textit{Middle}, the mean (points) and standard deviation (lines) in [Fe/H] for stars formed during each starburst, with the \texttt{NuGrid} runs showing substantially greater intrinsic scatter; in all cases, however, this decreases with time as turbulent diffusion mixes metals. \textit{Below}, non-normalized [Fe/H] MDF of our simulations at $z=0$. In the lower two panels, grey areas indicate the region of the ``metallicity floor'' ([Fe/H]$\sim-4$) in our simulations, where the earliest-formed stars artificially accumulate. 
    }
    \label{fig:sfh}
\end{figure}
\begin{figure}
    \centering
    \includegraphics[width=0.469\textwidth]{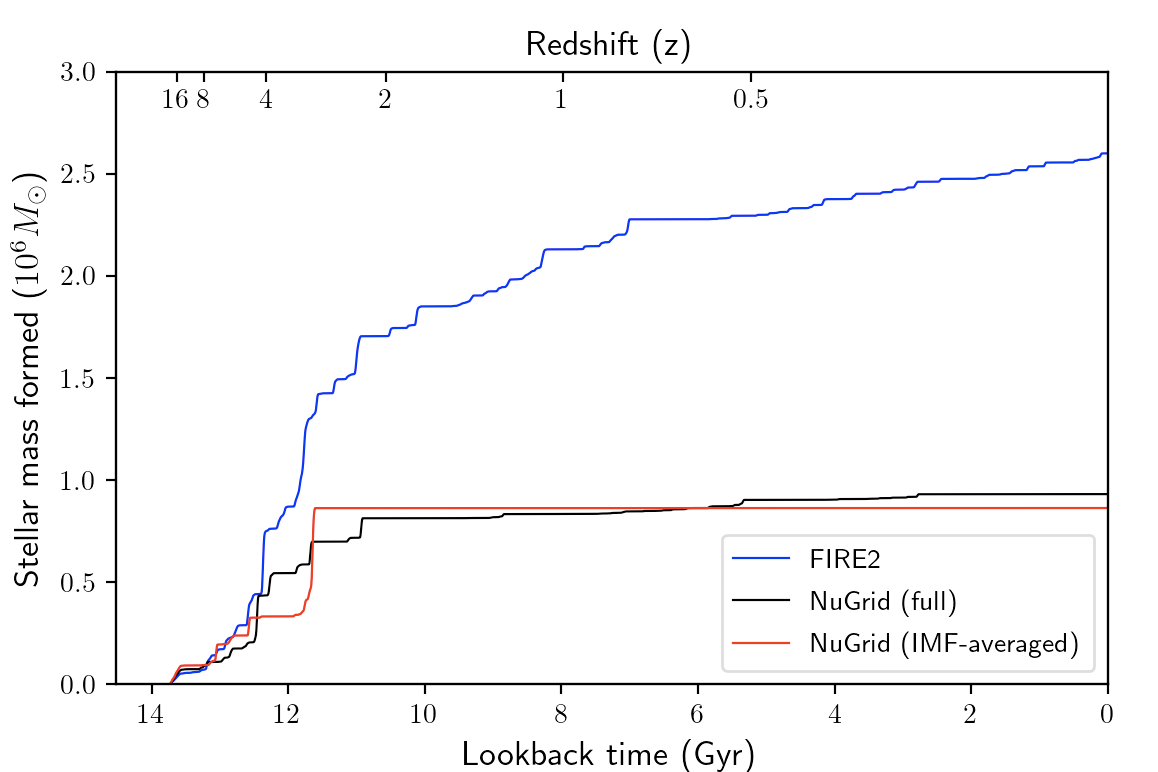}
    \caption{Cumulative star formation (in $M_\odot$) for our three runs, as a function of time. As demonstrated in Figure \ref{fig:variation_cumsf}, there is no clear association between SFH and the yield model used, and total stellar mass formed agrees to within a factor of ${\sim}3$ between all runs.}
    \label{fig:sfh_cumulative}
\end{figure}

We plot the star-formation rates (SFRs) as a function of lookback time for the three simulated galaxies in the upper panel of Figure \ref{fig:sfh}; in all cases, the SFR is bursty, peaking at $z \sim 4$ and declining thereafter. In each case, the cumulative star-formation history is available in \ref{fig:sfh_cumulative}. Both \nugridrun \ and \nugridavgrun \  produce ${\sim} 10^6 M_\odot$ of stars over their lifetime, with \nugridavgrun~quenching its star formation at $z \approx 2$. \defaultfirerun, however, forms ${\sim}2.7 \times 10^6 M_\odot$, with ${\sim}30\%$ of that mass being formed after $z = 2$. The differences in star formation histories between runs are consistent with stochastic run-to-run variation that has been seen in other investigations of this particular galaxy \citep{ElBadry2017,Hopkins2018,Su2018,Keller2019,Wheeler19}, due to the effects of random seeds and non-deterministic MPI reduction. We demonstrate this further in Figure \ref{fig:variation_cumsf}, where we show the cumulative fractional star formation histories (SFHs) for all three runs. The yield model has no consistent effect on star formation history or total stellar mass formed. We thus conclude that the choice between FIRE-2-default or \texttt{NuGrid} feedback parameters (supernova rate, energy, and ejecta mass), as well as including progenitor mass-dependence, does not have a qualitatively significant impact on the dynamics and star formation of the simulated galaxies.

\subsection{Metallicity distributions}
\label{sec:chem_evo}

\begin{figure*}
    \centering
    \includegraphics[width=0.95\textwidth,trim={0 2.5 0 0},clip]{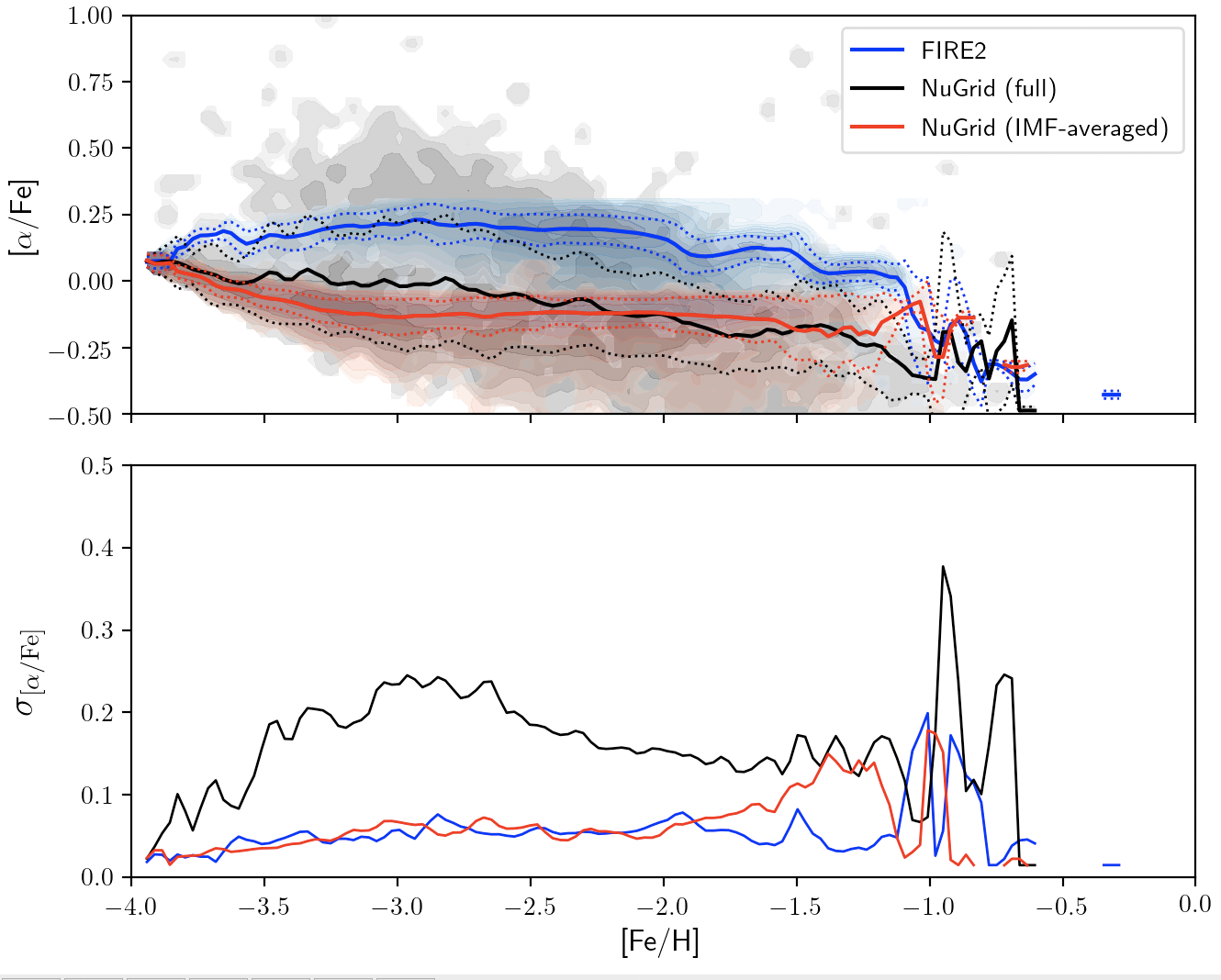}
    \caption{\textit{Above}, probability densities in [$\alpha$/Fe] vs. [Fe/H]-space for star particles in our simulations (\defaultfirerun \ \textit{blue}, \nugridrun \ \textit{black/grey}, \nugridavgrun \  \textit{red}). Solid lines indicate average [$\alpha$/Fe] at constant [Fe/H], while dotted lines indicate the $\pm 1 \sigma$ intrinsic-scatter interval. \textit{Below}, a more detailed plot of $1 \sigma$ intrinsic scatter; the $\sigma \approx 0.1-0.3$ dex we find for \nugridrun \ is indicated in some observations (see main text).}
    \label{fig:alpha-fe}
\end{figure*}
In the lower panel of Figure \ref{fig:sfh}, we plot the $z=0$ stellar-mass-weighted metallicity distribution function (MDF) for [Fe/H] in each of our simulations. All MDFs peak at an ${\rm [Fe/H]} \approx -2.2$, approximately in line with the observed MDF of Ursa Minor, due to the inclusion of metal diffusion in the runs \citep{Escala18}. However, the average metallicities \citep[in log space; see][for a definition]{Escala18} are lower, and standard deviations higher, in \nugridrun \ and \nugridavgrun \ than in \defaultfirerun, with $\left<{\rm [Fe/H]}\right>_{\rm FIRE-2} = -2.1 \pm 0.5$, while $\left<{\rm [Fe/H]}\right>_{\rm NuGrid  \ full} = -2.5 \pm 0.7$ and $\left<{\rm [Fe/H]}\right>_{\rm NuGrid \ avg} = -2.6 \pm 0.7$; these figures reflect both the lower level of star formation in the \texttt{NuGrid} runs, and the fatter low-metallicity tails of the \texttt{NuGrid} MDFs, including many stars at the metallicity ``floor'' of the simulation ([Fe/H]$\sim-4$). Physically, the tails result from stochastic spatial and temporal variation in star formation---because metals are not yet well-mixed by diffusion at early times, there is an increased scatter in [Fe/H] within and between the initial episodes of star-formation, whose magnitude can differ from run to run.

To better understand metal enrichment, we also study the mean and dispersion in metallicity for individual starbursts (defined when star formation crosses $1 M_\odot/{\rm Myr}$) in the middle panel of Figure \ref{fig:sfh}. We find that the typical iron abundance of each starburst, $\left<{\rm [Fe/H]}\right>_{\rm sb}$, increases with time for all simulations, although it is lower by roughly 0.17 dex between \nugridrun \ and \defaultfirerun \ due to lower star formation in the former.\footnote{The \nugridavgrun \  run does not show starbursts at late times with which to make a comparison, but early on follows the same pattern as the \nugridrun.} For starbursts at lookback times less than 10 Gyr (equivalently, $z \lesssim 2$), we find that $\sigma_{\rm [Fe/H], sb} \approx 0.05 {\rm \ dex}$ for simulations with default FIRE-2 yields, and $\sim 0.1-0.2 {\rm \ dex}$ for the full-\texttt{NuGrid} simulation; this intrinsic scatter decreases with time in all simulations \citep{Kirby13, Escala18} as turbulent diffusion brings metal concentrations into equilibrium. There is a positive correlation between $\sigma_{\rm [Fe/H], sb}$ and burst duration for \nugridrun, which is absent in the other (IMF-averaged) runs; this suggests that short-term (e.g., time-dependent SNe II metal yields), long-term (e.g., Type Ia and AGB enrichment), and numerical factors (turbulent diffusion, stochastic star formation) all contribute to the measured intrinsic scatter. To leading order, however, the width and shape of the MDF is predominated by differences in metallicity \textit{between} stellar populations formed at different times.


\subsection{Alpha-element abundances}
\label{ref:alpha_element_sec}
In the upper panel of Figure \ref{fig:alpha-fe}, we show the full distribution (shaded regions), mean value (solid lines) and $1{\rm -}\sigma$ scatter (dotted lines) of [$\alpha$/Fe] versus [Fe/H] for stars in all our simulations. We find, on average, that the mean abundance patterns for \nugridrun \ and \nugridavgrun \ agree with one another, particularly at high metallicity. For both \defaultfirerun \ and \nugridavgrun, we find that the trend is very narrow, with a scatter of only $\sigma_{{\rm [}\alpha{\rm / Fe]}} \approx 0.05$ dex throughout the metallicity range we test for both stars and gas. With the fully time-dependent yields in \nugridrun, however, the scatter rises substantially to $\sigma_{{\rm [}\alpha{\rm / Fe]}} \approx 0.1-0.3$ dex\footnote{Note that the sharp spike in $\sigma_{{\rm [}\alpha{\rm / Fe]}}$ for this run near [Fe/H] $\sim -1$ is an artifact of the low-number of stars at that metallicity.}. This occurs because, with full-\texttt{NuGrid} yields, a star particle at a given [Fe/H] can produce supernovae with \textit{multiple} possible [$\alpha$/Fe] abundances, as opposed to just one with IMF-averaged yields. 

Not only are non-IMF-averaged yields more conceptually realistic, they also better reflect the [$\alpha$/Fe]-[Fe/H] features observed in some dwarf galaxies. While a detailed comparison of intrinsic scatter between theory and observation is not appropriate due to the widely varying methods for determining systematic error between samples, we can comment on the broad trends seen in some observations that would likely be better matched by mass- and metallicity-dependent yields. \citet{Aoki2009} observe a sample of six extremely metal-poor stars in Sextans, finding for the first time, a star with low [$\alpha$/Fe] at low [Fe/H], hinting at a larger scatter in alpha abundances than is found in the halo of the Milky Way. This can also be seen in Sculptor and Carina \citep{Kirby09,Lemasle2012}.
Additionally, there is evidence that the M31 dwarf spheroidal galaxies may have more pronounced scatter than dSph satellites of the MW \citep{Vargas2014,Kirby19}. Using a method similar to that outlined in \citet{Escala18}, we find the intrinsic scatter in [alpha/Fe] for the five dSph satellites presented in \citet{Kirby19} to be 0.1-0.3 dex, while scatter in [alpha/Fe] for the MW satellites from \citet{Kirby11} is consistent with being entirely due to observational error. While we note that the error found for M31 satellites is similar to the scatter seen in \nugridrun, we caution that the method of estimating systematic error may add significant uncertainty to these results (Kirby, private communication).

\nugridrun~ also exhibits an increasingly pronounced intrinsic scatter for lower-metallicity stars. This trend is difficult to observe in dwarf galaxies, due to the low number of stars observed, but is seen in observations of halo stars \citep{Ryan1996}, and is at least hinted at in some dwarf galaxy observations at low metallicity (\citealt{Vargas2013,Vargas2014}; see also anomalous stars referenced in \citealt{Frebel15}). We note that \nugridrun~ broadly matches the observed trend of decreasing $\rm [\alpha/Fe]$ with increasing [Fe/H], although flat trends have also been observed in some galaxies \citep{Aoki2009, Frebel2010, Shetrone2001, Tafelmeyer2010}.

\citet{Revaz2016} show that scatter in [$\alpha$/Fe] vs [Fe/H] (specifically [Mg/Fe]) is artificially increased at sufficiently high resolution $(m_{\rm bar} \lesssim 1000 \msun$) in simulations when stochastic sampling of the IMF and mass-dependent yields are used. They show that the scatter can be brought back down to observed levels ($\sigma \lesssim 0.3$ dex for an ensemble of observed dwarf galaxies), by including either a metal mixing or a metal smoothing scheme. All of our runs include turbulent metal mixing, have high mass resolution, and all have $\sigma \lesssim 0.3$. However, the lower panel of \ref{fig:alpha-fe} shows a striking difference in the scatter in [$\alpha$/Fe] at fixed [Fe/H] between \defaultfirerun~and \nugridavgrun~when compared to \nugridrun. This difference in scatter is also seen when using a second set of runs for each yield model (see Appendix A), demonstrating that variations in the IMF have a much larger impact on scatter in [$\alpha$/Fe] vs [Fe/H] than does run-to-run variation. This is important, as run-to-run variation can be seen to drastically affect the star formation history and stellar mass of a given run (see Figure \ref{fig:variation_cumsf}). This means that, for a given star born in a dwarf galaxy, when and under what galactic conditions it was born may have less of an effect on its abundances than the mass of stars that exploded in its neighborhood prior to its birth.

In \texttt{NuGrid}, Type II core-collapse supernova ejecta from the most massive progenitors (with lifetimes ${\lesssim}8$ Myr, as seen in Figure \ref{fig:sb99_nugrid_full}) exhibit a high ratio of [$\alpha$/Fe] despite low overall production of $\alpha$-elements and Fe. Ejecta from lower-mass progenitors, by contrast, contain substantially more Fe and therefore a lower [$\alpha$/Fe] ratio. In \nugridrun, both types of supernovae enrich gas, and (through star formation that occurs more rapidly than turbulent diffusion eliminates [$\alpha$/Fe] gradients) become represented on the [$\alpha$/Fe] sequence. In \nugridavgrun, however, IMF-averaging means that Fe-poor (high-mass) yields are pre-mixed with the Fe-rich (low-mass) yields, and so have little effect on the final [$\alpha$/Fe] of the ejecta. Between $-3.5 \lesssim {\rm [Fe/H]} \lesssim -2.5$, this effect causes [$\alpha$/Fe] to be systematically ${\sim}0.1$ dex higher in \nugridrun \ than in \nugridavgrun.

More substantial is the difference in $\alpha$-abundances between the \defaultfirerun \ and \nugridavgrun \ simulations, with $\left<{\rm [}\alpha/{\rm Fe ]}\right>_{\rm FIRE-2} - \left<{\rm [}\alpha/{\rm Fe ]}\right>_{\rm NuGrid \ avg} \approx 0.2-0.4 {\rm \ dex}$, depending on [Fe/H]---in agreement with the expected IMF-averaged difference in [$\alpha$/Fe] production per core-collapse supernova between \cite{Nomoto2006} and \texttt{NuGrid} (see Sec. \ref{sec:typeII}). 
Observations suggest that for most stars in most dwarf galaxies, [$\alpha$/Fe] at low [Fe/H] is similar to the abundance of the Galactic halo \citep{Frebel2010, Hill2019, Kirby2010, Tolstoy09}. However, there are some instances of stars having [$\alpha$/Fe] near or even below solar at extremely low [Fe/H] \citep{Shetrone2001,Aoki2009, Tafelmeyer2010, Vargas2013, Starkenburg2013}, more closely matching the abundances in \nugridrun. Of course this could be high scatter masquerading as a lower normalization, but the low number of stars at low [Fe/H] in dwarfs makes this difficult to determine. We conclude that, despite capturing a larger \textit{dispersion} in [$\alpha$/Fe] versus [Fe/H], and matching \textit{some} observed low values of [$\alpha$/Fe] at low [Fe/H], the \texttt{NuGrid} tables may struggle to reproduce the observed $\alpha$ abundance values at low metallicity found in most dwarf galaxies.

\section{Conclusions}
\label{sec:conclusion}

We present the a detailed case-study of cosmological hydrodynamic simulations run with different SNe properties (energy, mass, and yields) by comparing a single dwarf galaxy run with the default FIRE-2 models \citep[\defaultfirerun;][]{Hopkins2018} to a more detailed model which allows for complicated progenitor-mass and metallicity dependence of all of the above from \texttt{NuGrid} (\nugridrun). We also include a run in which we use the yields from \texttt{NuGrid}, but average over the IMF (\nugridavgrun).

We find that allowing for variations in the mass, metallicity, and energy of individual explosions has only a weak effect on integral galaxy properties (e.g. stellar masses, SFRs, total metallicities, MZR, etc.), well within the range of stochastic variations. This is not surprising, as these properties are primarily sensitive to other integral quantities (e.g.\ total energy of SNe), as shown in the detailed study of \citet{Su2018} who considered much more radical explosion-to-explosion variations (e.g.\ hypernovae with $\sim1000x$ higher energy, as opposed to factor $<2$ variations in SNe energy in the \texttt{NuGrid} models). 

As expected, the spread in detailed abundances or [$\alpha$/Fe] ratios at a given time -- or equivalently, value of [Fe/H] -- is larger when we include progenitor mass-and-metallicity dependent SNe yields. Moreover, with the more detailed models, the scatter in [$\alpha$/Fe] {\em increases} towards lower metallicity, in a manner similar to that observed in some dwarf galaxies, as the enrichment becomes more sensitive to the effect of individual SNe (as compared to a well-mixed system which has been enriched by many SNe, and therefore reflects an IMF-averaged population). For the systems here, above [Fe/H] ${\gtrsim}-1$, IMF-averaging appears reasonable in the detailed abundance tracks, while below this, IMF averaging likely produces too little variation in individual stellar abundance patterns. Thus, for future studies of metal-poor stellar populations, it will be important to include more detailed enrichment models of the sort studied here.

However, although the width and shape of the [$\alpha$/Fe] distribution and MDF in our detailed \texttt{NuGrid} models agrees reasonably well with observations, the absolute value of [$\alpha$/Fe] is systematically lower than that observed in \textit{most} dwarf galaxies.
This owes simply to the fact that even {\em pure} core-collapse SNe yields (at {\em any} progenitor metallicity $Z<-1$), in IMF-average, give sub-solar [$\alpha$/Fe] in the \texttt{NuGrid} models. This is primarily driven by a combination of relatively high Fe yields for core-collapse explosions, and much lower (factor $\sim3-5$) Mg, compared to other models such as those in \citet{Nomoto2006} as adopted in FIRE-2. If future observations of  [$\alpha$/Fe] in extremely low metallicity stars continue to suggest a normalization $\sim 0.3-0.6$\,dex higher than \nugridrun~ and \nugridavgrun, it is possible that some of \texttt{NuGrid}'s nucleosynthetic yields may need to be recalibrated.

\section*{Acknowledgements}
We thank Benoit C{\^o}t{\'e}, Andrew Graus, Alexander Ji, Evan Kirby, Mariska Kriek, Shea Garrison-Kimmel, Michael Grudi{\'c}, Andrew Graus, Xiangcheng Ma, and Kung-Yi Su for useful discussions. Support for DM and coauthors was provided by NSF Collaborative Research Grants 1715847 \&\ 1911233, NSF CAREER grant 1455342, NASA grants 80NSSC18K0562, JPL 1589742. Numerical calculations were run on the Caltech compute cluster ``Wheeler,'' allocations FTA-Hopkins supported by the NSF and TACC, and NASA HEC SMD-16-7592. The data used in this work were, in part, hosted on facilities supported by the Scientific Computing Core at the Flatiron Institute, a division of the Simons Foundation. CRW acknowledges support from NASA through the NASA Hubble Fellowship grant \#HST-HF2-51449.001-A, awarded by the Space Telescope Science Institute, which is operated by the Association of Universities for Research in Astronomy, Inc., for NASA, under contract NAS5-26555.  AW received support from NASA through ATP grant 80NSSC18K1097 and HST grants GO-14734, AR-15057, AR-15809, and GO-15902 from STScI; a Scialog Award from the Heising-Simons Foundation; and a Hellman Fellowship. DK was supported by NSF Grant AST-1715101 and by a Cottrell Scholar Award from the Research Corporation for Science Advancement.


\section*{Data availability}
The data underlying this work will be shared at reasonable request to the corresponding author.




\bibliographystyle{mnras}




\appendix
\section{Run-to-run variation}
\label{sec:appendix_a}

\begin{figure}
    \centering
    \includegraphics[width=0.495\textwidth]{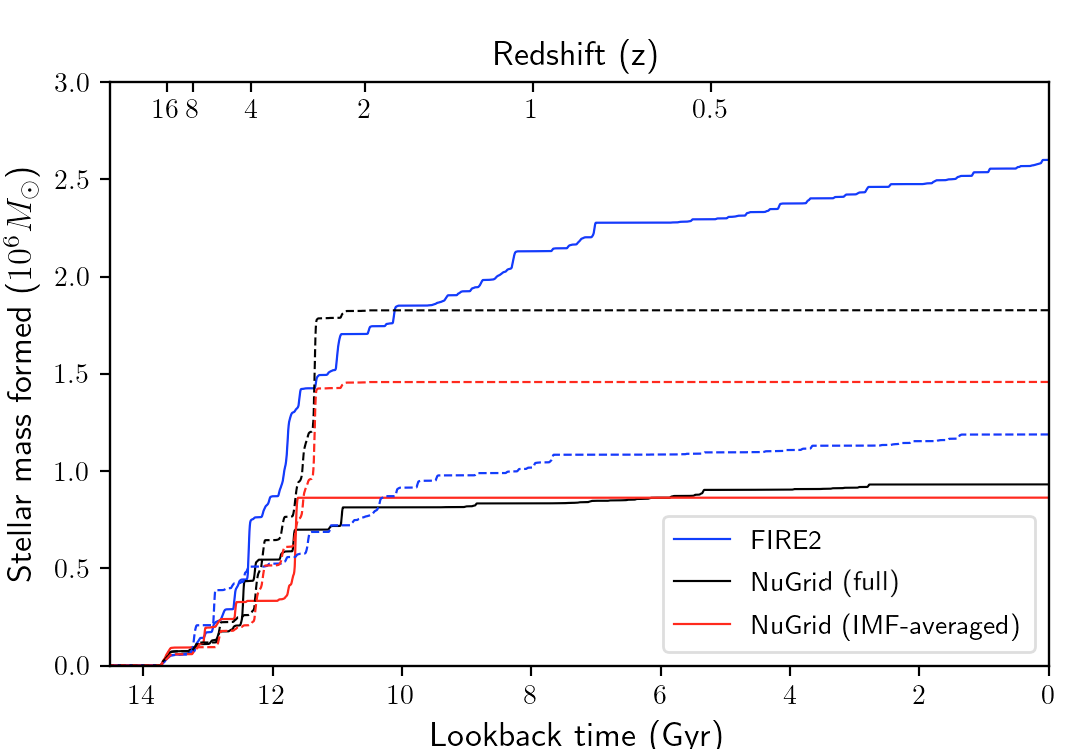}
    \caption{Cumulative star-formation histories for the  \defaultfirerun, \nugridrun, and \nugridavgrun \ runs \textit{(solid)} discussed in the text, as well as the corresponding replication runs \textit{(dashed)}. There is no clear association between SFH and the yield model used, and total stellar mass formed agrees to within a factor of ${\sim}3$ between all runs.}
    \label{fig:variation_cumsf}
\end{figure}

\begin{figure}
    \centering
    \includegraphics[width=0.495\textwidth]{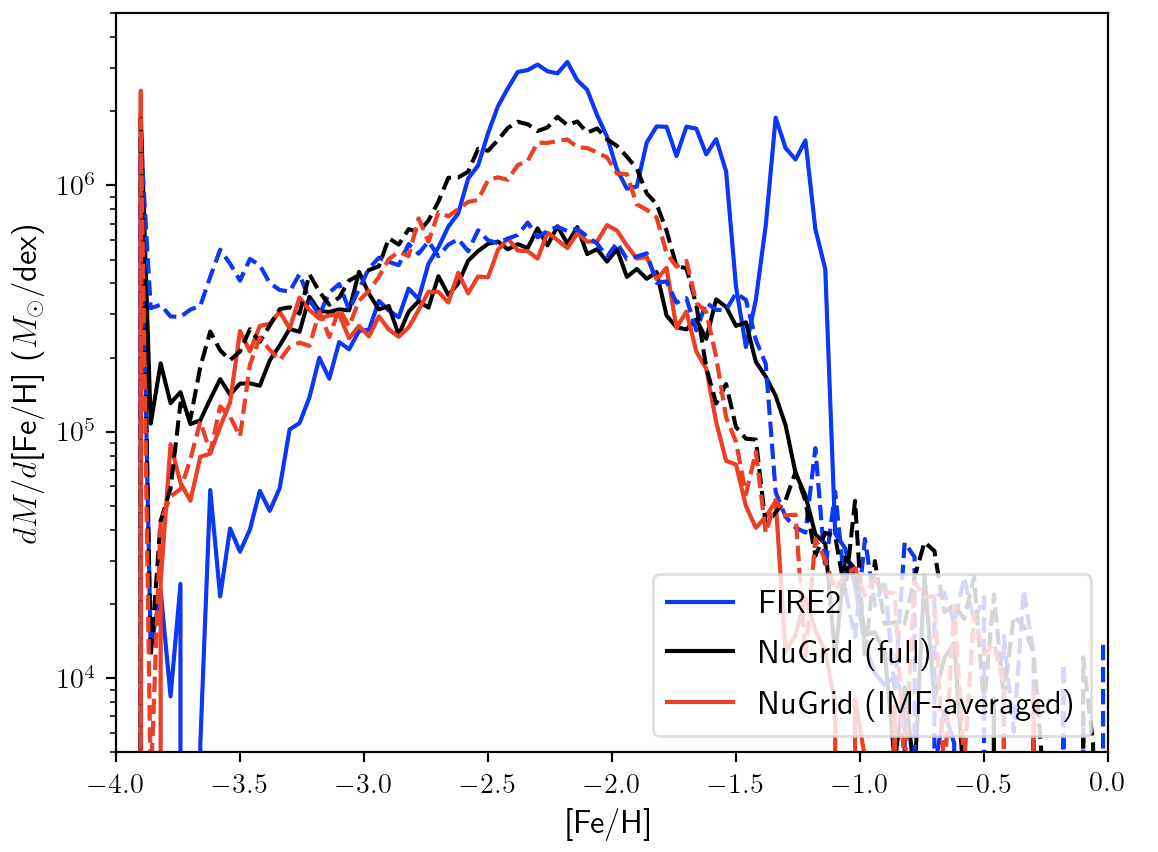}
    \caption{[Fe/H] distributions for our fiducial \textit{(solid)} and replication \textit{(dashed)} simulations. Differences between runs, particularly the low-metallicity tail, are driven by stochastic variation in star-formation history, and show little correlation with the enrichment prescription used. Above [Fe/H] $\approx -1$, there is little star formation, leading to substantial sampling error in the distribution.}
    \label{fig:fe_h_sequence_replication}
\end{figure}

\begin{figure}
    \centering
    \includegraphics[width=0.495\textwidth]{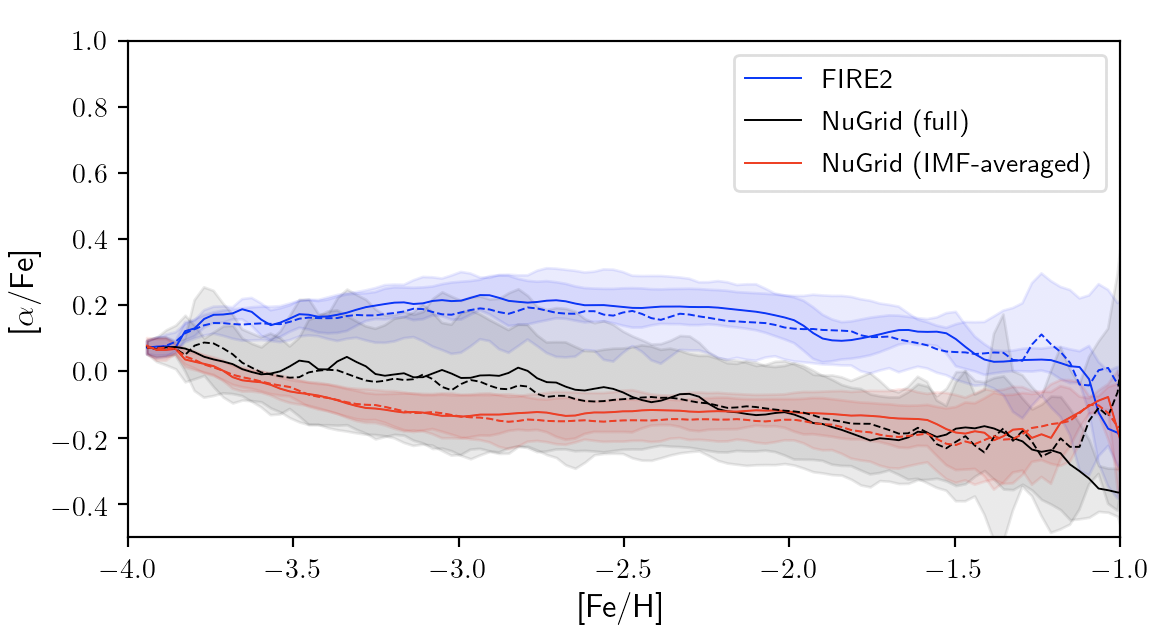}
    \caption{[$\alpha$/Fe] versus [Fe/H] sequences for our fiducial \textit{(solid)} and replication \textit{(dashed)} simulations, with 1-sigma confidence intervals shaded; above [Fe/H] $\approx -1$, sampling error means the sequences are poorly resolved. Despite substantial run-to-run variation in star-formation history, the sequences are well-converged, and demonstrate that time-dependent yields contribute to a greater intrinsic scatter.}
    \label{fig:a_fe_sequence_replication}
\end{figure}

As can be seen in Figure \ref{fig:sfh}, the duration of star formation varies substantially between the \defaultfirerun, \nugridrun, and \nugridavgrun~runs. We argue in Section \ref{sec:sfh} that these differences are consistent with stochastic run-to-run variation between simulations, and are not related to the yield model used. In Figure \ref{fig:variation_cumsf}, we plot cumulative fractional star formation histories (SFHs) for the original three runs (solid lines) alongside an additional run of each model with identical physical initial conditions (dotted lines). The stellar mass at $z=0$ for each run is consistent to within a factor of ${\sim}3$, with no clear relationship to the yield model employed. 

In \ref{fig:fe_h_sequence_replication}, we plot [Fe/H] distribution fucntions for our fiducial and replication runs. There is no clear dependence on the metal-enrichment prescription, which strongly suggests that stochastic variations in final stellar mass and quenching time drive the differences between runs. Despite this, however, we show in Figure \ref{fig:a_fe_sequence_replication} that the [$\alpha$/Fe] versus [Fe/H] tracks are nearly identical for each run of a given yield model. This strongly suggests that the conclusions presented here are robust against stochastic run-to-run variation. Variations in the IMF have a larger effect on scatter than variations in the star formation history of the galaxy.


\bsp	
\label{lastpage}
\end{document}